\documentclass[12pt]{article}
\usepackage{graphics}

\usepackage{color}
\usepackage{epstopdf}
\usepackage{amsmath}
\usepackage{amssymb}
\usepackage{booktabs}
\usepackage{graphicx}
\usepackage{xcolor,colortbl}
\usepackage{lipsum}
\usepackage{setspace}
\usepackage[numbers]{natbib}
\usepackage{multirow}
\usepackage{bm}
\usepackage{enumitem}
\usepackage{comment}
\usepackage{soul}
\usepackage{hyperref}

\newcommand {\rb}[1]{\raisebox{1.5ex}[0pt]{#1}}

\begin{document}

\title{\large Return-to-baseline multiple imputation for missing values in clinical trials}
\author{
{\bf\small Yongming Qu*, Biyue Dai} \\ \small Department of Statistics, Data and Analytics, Eli Lilly and Company, Indianapolis, IN 46285, USA \\
}
\date{\small \em \today}
\maketitle
\noindent
{\small *Correspondence: Yongming Qu, Department of Statistics, Data and Analytics, Eli Lilly and Company, Lilly Corporate Center, Indianapolis, IN 46285, USA. Email: qu\_yongming@lilly.com.}

\begin{abstract}
Return-to-baseline is an important method to impute missing values or unobserved potential outcomes when certain hypothetical strategies are used to handle intercurrent events in clinical trials. Current return-to-baseline approaches seen in literature and in practice inflate the variability of the ``complete" dataset after imputation and lead to biased mean estimators {when the probability of missingness depends on the observed baseline and/or postbaseline intermediate outcomes}. In this article, we first provide a set of criteria a return-to-baseline imputation method should satisfy. Under this framework, we propose a novel return-to-baseline imputation method. Simulations show the completed data after the new imputation approach have the proper distribution, and the estimators based on the new imputation method outperform the traditional method in terms of both bias and variance, when missingness depends on the observed values. The new method can be implemented easily with the existing multiple imputation procedures in commonly used statistical packages.
\\
{\textbf{KEY WORDS}: Baseline observation carried forward; direct maximum likelihood estimation; estimand; ignorable missingness. }
\end{abstract}

\newpage 

\section{Introduction} \label{sec:introduction}
{
The ICH E9(R1) addendum \cite{ich2019e9} provides a general framework for estimands and sensitivity analyses for clinical trials. Missing values can be truly missing measurements, unobserved outcomes, or discarded observations as a result of using a hypothetical strategy handling intercurrent events.
For example, a patient discontinues the study medication due to an adverse event and subsequently takes an alternative medication. Sometimes we may consider this patient to have no benefit from the study medication, as the patient cannot tolerate the medication \citep{qu2020defining,qu2021implementation}. This patient either discontinues from the study and does not come back to have the primary outcome measured, or completes the study with a measurement for the primary outcome which does not reflect the potential outcome of interest. In either situation, it results in a missing value. {\color{black}
If the potential outcome of interest is the treatment effect for patients who have not used additional rescue medications, the missingness in potential outcome resulting from patients' early discontinuation of treatment may be imputed using \emph{retrieved dropouts} \cite{CHMP2010}, i.e., patients who discontinue treatment without taking rescue medications and with observed primary outcomes.} Alternatively, in a placebo-controlled study, one may use a reference-based imputation method (e.g, \emph{jump-to-reference} imputation \cite{carpenter2013analysis}). {For diseases with slow progression and for treatments without disease modification, when there are not enough cases of retrieved dropouts in an active-comparator study and placebo effect is expected to be low, \emph{return-to-baseline} imputation, which assumes the potential outcome has the same mean as baseline, may be used as an alternative.} In addition, the return-to-baseline imputation may also be used for sensitivity analyses, assuming the missing outcomes have values that are as poor as baseline (no improvement) for certain situations.} {Therefore, it is important to have a robust return-to-baseline imputation method that can serve as one of many tools used in certain clinical trial data analyses.}

One simple but naive approach to impute the missing values under such {\color{black}a strategy} is the baseline observation carried forward (BOCF) method. While BOCF is simple and intuitive, it ignores the within-subject variability of the measurements at different time points. {Therefore, BOCF will likely strengthen the correlation between measurements at baseline and endpoint, and hence weaken the correlation between baseline value and the change from baseline.} The missing value under the return-to-baseline assumption can also be handled by the direct likelihood-based method \citep{zhang2020likelihood}, but it lacks flexibility when missing values may need to be imputed differently based on reason of missingness \citep{qu2020revisiting}. For example, missing values simply due to an invalid measurement may be imputed differently than missing values due to intercurrent events.

Alternatively, {a more appropriate return-to-baseline imputation procedure should be based on the principal of multiple imputation \cite{rubin1976inference}, in which} the imputed values should have the same mean as the baseline values, but with {uncertainties incorporated through a random error}. Although the idea is simple, implementation is not uniformly agreed upon and several versions of implementation are available. Quan et al.  \cite{quan2018considerations} propose using a bivariate normal distribution between the baseline value and the potential outcome of the post-baseline value, assuming the potential outcome and the baseline value have the same mean and variance. The imputed values produced by this approach will have the same distribution as the baseline values. However, the correlation between baseline and post-baseline values cannot be estimated from the observed data. The authors suggest an approximation, but the variance of imputed values is much larger than the variance of the baseline values. The approximation will also result in the imputed mean being different from the baseline mean {when the missingness depends on observed baseline and/or postbaseline intermediate outcomes (MDO)}. Zhang et al. \cite{zhang2020missing} propose a return-to-baseline imputation by adding additional random error to the baseline value. {This approach shares the same drawbacks as the approximation introduced by Quan et al. \cite{quan2018considerations}. }

In this article, we propose a new return-to-baseline imputation valid {when the probability of missingness can be modeled through observed outcomes}. Section 2 conducts a brief review of the existing methods, and proposes a general {\color{black}theoretical} framework and the new method for the return-to-baseline imputation. Section 3 details the simulation experiments conducted to evaluate the empirical performance of the new method. Finally, Section 4 summarizes the key findings, followed by a discussion.

\section{Methods} \label{sec:methods}
Assume we have $I$ randomized treatment groups indexed by $i=1,2,\ldots, I$ and let $n_i$ denote the sample size for group $i$. Let $Y_{ij}$ denote the outcome variable (typically measured at the end of the study treatment period) and $X_{ij}$ denote the baseline value of the outcome variable, for subject $j$ ($j=1,2,\ldots,n_i$) in treatment group $i$. Let $Z_{ij}$ denote the vector of the ancillary data (baseline or/and postbaseline) that may be potentially correlated with $Y_{ij}$. {The ancillary variable $Z_{ij}$ can include the measurements for the outcome variable $Y_{ij}$ at the intermediate time points and other ancillary variables either measured at baseline or at postbaseline time points.} Assume $(X_{ij}, Z_{ij}', Y_{ij})'$ are from a multivariate normal distribution
\[
(X_{ij}, Z_{ij}', Y_{ij})' \sim N(\mu_i, \Sigma_i),
\]
where $\mu_i = (\mu_x, \mu_{z,i}', \mu_{y,i})'$ and $\Sigma_i$ is the variance-covariance matrix of $(X_{ij}, Z_{ij}', Y_{ij})'$.

We first assume the baseline values $X_{ij}$ and ancillary variables $Z_{ij}$ are not missing, and the outcome variable $Y_{ij}$ may be missing. Let $R_{ij}$ be the missingness indicator ($R_{ij}=0$ indicates $Y_{ij}$ is missing). {We assume ignorable missingness for the outcome $Y_{ij}$ to allow the probability of missingness to depend on $X_{ij}$ and $Z_{ij}$}:
\begin{equation} \label{eq:MAR}
R_{ij} \perp Y_{ij} | (X_{ij}, Z_{ij}).
\end{equation}
Let the probability of non-missingness be modeled as
\[
\pi_{ij} := \pi_{i}(X_{ij}, Z_{ij}) = \Pr(R_{ij} = 1|X_{ij}, Z_{ij}).
\]

\subsection{Return-to-baseline multiple imputation} \label{sec:rtb_mi}
The return-to-baseline imputation involves imputing the missing value by a value similar to the baseline value while accounting for within-subject variability. 

To study the property of an imputation method, we first need to define the potential outcome to impute. Under the causal estimand framework \cite{lipkovich2020causal}, we let $Y_{ij}(i,b)$ denote the potential outcome for patient $j$ assigned to treatment $i$ but taking treatment $b$. When a subject adheres to the assigned treatment, the potential outcome is denoted by $Y_{ij}(i,i)$, simplified as $Y_{ij}$. When a subject experiences a certain type of intercurrent event assumed to cause no improvement in the outcome from baseline, the potential outcome is denoted by $Y_{ij}(i,*)$, simplified as $Y_{ij}^*$. {Note even with the missing  ignorability assumption (\ref{eq:MAR}) for $Y_{ij}$, the missingness is NOT ignorable for $Y_{ij}^*$ because $R_{ij} \not\perp Y_{ij}^*|X, Z$. As mentioned in Section \ref{sec:introduction}, when the placebo effect is expected to be low, for disease with slow progression and for treatments without disease modification, \emph{return-to-baseline} imputation, which assumes the potential outcome has the same mean as baseline, may be used to imputed $Y_{ij}^*$.}
{Let $\tilde Y_{ij} = R_{ij} Y_{ij} + (1-R_{ij}) Y_{ij}^*$ denote the complete data after imputation. In this article, we assume the primary analysis variable is the change from baseline in the primary outcome. Therefore, we will study the return-to-baseline imputation for the distribution of $Y_{ij}^*$ and $\tilde Y_{ij}$, and the inference for the change from baseline $\Delta \tilde Y_{ij} =\tilde Y_{ij} - X_{ij}$}.

Instead of imputing the missing value to be equal to $X_{ij}$ (BOCF), return-to-baseline imputation relies on the joint distribution of $X_{ij}$ and $Y_{ij}^*$ to impute the missing value in $Y_{ij}^*$. A plausible return-to-baseline imputation $Y_{ij}^*$ should at least meet the following 2 criteria:



\newcommand{\subscript}[2]{$#1 _ #2$}
\begin{enumerate}[label=\subscript{C}{{\arabic*}}.]
    \item \label{C1} $E[Y_{ij}^*] = \mu_x$.
    \item The distribution of the complete data (observed and imputed) of the response variable should be the same as baseline values if $Y_{ij}$ and $X_{ij}$ have the same distribution.
\end{enumerate}
C2 is very important because return-to-baseline imputation should provide a consistent estimator for the population mean when the true mean for the outcome variable $Y$ is equal to the true mean for the baseline variable $X$. {For example, if there is no disease progression from baseline to the end of the study, and with little placebo effect, we expect the mean response at the end of the study to be equal to the baseline mean. }

The key challenge in return-to-baseline imputation is the unknown correlation between $X_{ij}$ and $Y_{ij}^*$. 
Quan et al. \cite{quan2018considerations} proposed to impute the return-to-baseline value from the distribution of $N(\mu_x +\rho_{i}(X_{ij}-\mu_x), \sigma_x (1-\rho_i^2))$, which is the conditional distribution of {\color{black}$Y_{ij}^*|X_{ij}$ if $Y_{ij}^*$} and $X_{ij}$ have the same mean and standard deviation with a correlation of $\rho_i$. The challenge of this method is there is no way to estimate the correlation $\rho_i$. One approximation suggested by the authors is to generate $Y_{ij}^q \sim N(X_{ij}, 2\tau^2)$, where $\tau^2$ is the mean squared difference between baseline and the first postbaseline measurement. 
{This approximation will make the imputed data have much larger variance than $X$, which can be seen from $Var(Y_{ij}^q) = Var(X_{ij}) + 2\tau^2 > Var(X_{ij})$}. 

Zhang et al. \cite{zhang2020missing} suggest the return-to-baseline imputation be $Y_{ij}^z = X_{ij} + \eta_{ij}$, where $\eta_{ij} \sim N(0, \tau^2)$ {independent of $X_{ij}$} and $\tau^2$ is estimated by the standard deviation of the change from baseline ($Y_{ij}-X_{ij}$) for all observed data. This is also a conservative imputation because $Var(Y_{ij}^z) = Var(X_{ij}) + \tau^2 > Var(X_{ij})$, although it is generally less conservative compared to the approximation suggested by Quan et al. \cite{quan2018considerations}. In addition, the standard deviation estimated from the {\em complete} data is not a consistent estimator of the standard deviation of $Y_{ij}-X_{ij}$ because {$Var(Y_{ij}-X_{ij}|R_{ij}=1) \ne Var(Y_{ij}-X_{ij})$ unless the missingness is independent of $X_{ij}$}. Even if the true standard deviation were used, such a method would still produce a biased mean for the imputed distribution, because $E[E(Y_{ij}^z|X_{ij}, Z_{ij}, R_{ij}=1)] \ne E[E(Y_{ij}^z|X_{ij}, Z_{ij})] = E(X_{ij}) = \mu_x$ unless $R_{ij} \perp (X_{ij}, Z_{ij})$, the case of missing completely at random (MCAR).

Here we propose borrowing the correlation between $X_{ij}$ and $Y_{ij}$ in the imputation. For the variance of $Y_{ij}^*$, we can either assume it is equal to $Var(Y_{ij})$ (only ``return" the mean to baseline) or $Var(X_{ij})$ (also ``return" the variance to baseline). In the remainder of this article, we primarily focus on the former case, but our method can be applied to the latter case as well. 

When $\mu$ and $\Sigma$ are known and $Y_{ij}$ is missing, a natural way to impute $Y_{ij}$ under \textcolor{black}{the assumption that} for each treatment group,  $\{(X_{ij}, Z_{ij}, Y_{ij}): j=1,2,\ldots,n_i\}$ are independently identically distributed, is to draw random samples from the conditional distribution of $Y_{ij}|X_{ij}, Z_{ij}$: 
\begin{equation} \label{eq:hypothetical_imputation}
Y_{ij}|X_{ij}, Z_{ij} \sim N(\mu_{y,i} + \Sigma_{21,i} \Sigma_{11,i}^{-1} (X_{ij}-\mu_x, (Z_{ij}-\mu_{z,i})')', \sigma_{y,i}^2 - \Sigma_{21,i} \Sigma_{11,i}^{-1} \Sigma_{12,i}),
\end{equation}
where $\Sigma_{11,i}$ is the variance-covariance matrix for $(X_{ij}, Z_{ij}')'$, $\sigma_{y,i}^2$ is the variance for $Y_{ij}$, and $\Sigma_{12,i}$ and $\Sigma_{21,i}$ are from the partition of 
\[
\Sigma_{i} = \left[ \begin{array}{cc} \Sigma_{11,i} & \Sigma_{12,i} \\ \Sigma_{21,i} & \sigma_{y,i}^2 \end{array} \right].
\]

{Return-to-baseline} imputation can be implemented similarly as in (\ref{eq:hypothetical_imputation}) but replacing $\mu_{y,i}$ with $\mu_x$:
\begin{equation} \label{eq:return_to_baseline_imputation}
Y_{ij}^*|X_{ij}, Z_{ij} \sim N(\mu_x + \Sigma_{21,i} \Sigma_{11,i}^{-1} (X_{ij}-\mu_x, (Z_{ij}-\mu_{z,i})')', \sigma_{y,i}^2 - \Sigma_{21,i} \Sigma_{11,i}^{-1} \Sigma_{12,i}).
\end{equation}
Criterion C1  can be immediately verified by $E(Y_{ij}^*) = E[ E(Y_{ij}^*|X_{ij}, Z_{ij})] = \mu_x$. 
If the distributions of $X_{ij}$ and $Y_{ij}$ are the same [e.g., $E(Y_{ij})=E(X_{ij})$ and $Var(Y_{ij}) = Var(X_{ij})$], then the conditional distributions of $Y_{ij}$ and $Y_{ij}^*$ are the same, given $X_{ij}$ and $Z_{ij}$. Hence, under the ignorable missingness assumption for $Y_{ij}$, the mean and variance for the complete data $\tilde Y_{ij}$ for treatment group $i$ are: 
\begin{eqnarray} 
&& E\left\{ R_{ij} Y_{ij} + (1-R_{ij}) Y_{ij}^* \right\}
\nonumber \\ &=& 
E\left[ E\left\{ R_{ij} Y_{ij} + (1-R_{ij}) Y_{ij}^* | R_{ij}, X_{ij}, Z_{ij}\right\} \right]
\nonumber \\  &=& 
E\left\{ R_{ij} E\left(Y_{ij}|X_{ij}, Z_{ij}\right) + (1-R_{ij}) E\left(Y_{ij}^* | X_{ij}, Z_{ij}\right) \right\} 
\nonumber \\ &=& 
E\left\{ E\left(Y_{ij}|X_{ij}, Z_{ij}\right) \right\}
\nonumber \\ &=& 
E(Y_{ij}) = E(X_{ij})
\label{eq:E_hybrid}
\end{eqnarray}
and
\begin{eqnarray} 
&& Var\left\{ R_{ij} Y_{ij} + (1-R_{ij}) Y_{ij}^* \right\}
\nonumber \\ &=& 
Var\left[ E\left\{ R_{ij} Y_{ij} + (1-R_{ij}) Y_{ij}^* | R_{ij}, X_{ij}, Z_{ij} \right\} \right] + 
E\left[ Var\left\{ R_{ij} Y_{ij} + (1-R_{ij}) Y_{ij}^* | R_{ij}, X_{ij}, Z_{ij} \right\} \right]
\nonumber \\ &=& 
Var\left\{ E\left(Y_{ij}|X_{ij}, Z_{ij}\right) \right\} 
+ E \left\{ Var\left(Y_{ij}|X_{ij}, Z_{ij}\right) \right\}
\nonumber \\ &=& 
Var(Y_{ij}) = Var(X_{ij}).
\label{eq:Var_hybrid}
\end{eqnarray}
Therefore, Criterion C2 is satisfied.  

In practice, the parameters $\mu_{z,i}$ and $\Sigma_i$ are unknown, but they can be consistently estimated using the method of maximum likelihood even {when the probability of missingness for $Z_{ij}$ and $Y_{ij}$ depends on the observed values}. When imputing missing data using the conditional distribution (\ref{eq:return_to_baseline_imputation}), the variabilities in the estimated parameters also need to be taken into account. In spite of the straightforwardness of the theoretical framework, programming such a multiple imputation procedure could be tedious, especially when the dimension of $Z_{ij}$ is high. Fortunately, we can utilize the existing multiple imputation packages/procedures to implement the newly proposed return-to-baseline imputation easily:
\begin{enumerate}
\item Impute the missing values for the potential outcome of \textcolor{black}{$Y_{ij}$ for each treatment group separately under the ignorable missingness assumptions} (using all observed data in $X_{ij}$, $Z_{ij}$, and $Y_{ij}$ in each treatment as the source data) with a standard multiple imputation procedure (``PROC MI" in SAS or ``mice" package in R). Let $Y_{ij}^{(m)}$ denote the imputed value for subject $j$ in treatment group $i$ in the $m^{th}$ imputation.
\item For each treatment group $i$, let $\hat \mu_{y,i}^{(m)}$ be the sample mean of $R_{ij} Y_{ij} + (1-R_{ij}) Y_{ij}^{(m)}$, the complete data after imputation. 
\item {R}eturn-to-baseline imputation is given by
\begin{equation} \label{eq:rtb_mi}
Y_{ij}^{*(m)} = \title Y_{ij}^{(m)} - \hat \mu_{y,i}^{(m)} + \bar X_{\cdot \cdot},  
\end{equation}
where $\bar X_{\cdot \cdot}$ is the sample mean of $X_{ij}$ across treatment groups (assuming baseline means between treatment groups are the same in randomized studies).
\end{enumerate}


{In all multiple imputation packages, the imputed values $Y_{ij}^{(m)}$ for $Y_{ij}$ under the \textcolor{black}{ignorable missingness} assumption are the random numbers generated from the conditional distribution in Equation (\ref{eq:hypothetical_imputation}) with appropriate uncertainty incorporated in the estimated parameters or through a different but equivalent process (e.g., Markov Chain Monte Carlo method). Therefore, Equation (\ref{eq:rtb_mi}), which adjusts for the mean difference between $Y_{ij}$ and $X_{ij}$, produces imputed values equivalent to random numbers directly generated from Equation (\ref{eq:return_to_baseline_imputation}), with the incorporation of the uncertainties in the {\color{black}estimated} parameters. 
}

{Imputation based on Equation (\ref{eq:rtb_mi}) only ``return" the mean to baseline.} If one is interested in obtaining imputed values that have the same variance as the variance of baseline values, Equation (\ref{eq:rtb_mi}) can be replaced with
\begin{equation} \label{eq:rtb_imputation2}
Y_{ij}^{*(m)} = S_x \left\{S_{y,i}^{(m)} \right\}^{-1} (\title Y_{ij}^{(m)} - \hat \mu_{y,i}^{(m)}) + \bar X_{\cdot\cdot},
\end{equation}
where $S_x$ is the sample standard deviation for $X_{ij}$ across treatment groups and $S_{y,i}^{(m)}$ is the sample variance for \textcolor{black}{$\{R_{ij} Y_{ij} + (1-R_{ij}) Y_{ij}^{(m)}\}_{j=1}^{n_i}$}. {For the remainder of this article we will only discuss the return-to-baseline imputation based on Equation (\ref{eq:rtb_mi}). }

After imputation, change from baseline can be calculated and the corresponding analyses can be performed. {The SAS and R program to implement the proposed return-to-baseline imputation is provided in the Appendix.}

\subsection{Implicit return-to-baseline imputation via direct (restricted) maximum likelihood estimation}
Let the complete data after the return-to-baseline imputation be $\tilde Y_{ij} = R_{ij} Y_{ij} + (1-R_{ij}) Y_{ij}^*$ and the corresponding change from baseline be  $\Delta \tilde Y_{ij} = \tilde Y_{ij} - X_{ij}$. Under the ignorable missingness assumption (\ref{eq:MAR}), the mean of $\Delta \tilde Y_{ij}$ can be expressed as
\begin{eqnarray} 
\mu_{\Delta \tilde y, i} &=&
E\left( \Delta \tilde Y_{ij}  \right)  \nonumber\\
&=& 
 E\left\{ R_{ij} Y_{ij} + (1-R_{ij}) Y_{ij}^* - X_{ij}\right\} 
\nonumber \\ &=& 
E\left[ E\left\{ R_{ij} Y_{ij} + (1-R_{ij}) Y_{ij}^* | R_{ij}, X_{ij}, W_{ij}\right\} \right] -\mu_x
\nonumber \\ &=& 
E\left\{ R_{ij} E\left(Y_{ij}|X_{ij}, W_{ij}\right) + (1-R_{ij}) E\left(Y_{ij}^* | X_{ij}, W_{ij}\right) \right\} -\mu_x
\nonumber \\ &=& 
E\left[ R_{ij} E\left(Y_{ij}|X_{ij}, W_{ij}\right) + (1-R_{ij}) \{E\left(Y_{ij} | X_{ij}, W_{ij}\right) - \mu_{y,i} + \mu_x \} \right] -\mu_x 
\nonumber \\ &=& 
E\left\{ E\left(Y_{ij}|X_{ij}, W_{ij}\right) - (1-R_{ij})(\mu_{y,i}-\mu_x) \right\} -\mu_x
\nonumber \\ &=& 
\mu_{y,i} - (1-\pi_{i})(\mu_{y,i}-\mu_x) - \mu_x
\nonumber \\ &=& 
\pi_{i} (\mu_{y,i}-\mu_x) 
\nonumber \\ &=& 
\pi_{i} \cdot \mu_{\Delta y,i},
\label{eq:E_ML}
\end{eqnarray}
where $\pi_i = \Pr(R_{ij}=1)$ and $\mu_{\Delta y,i} = E(\Delta Y_{ij})$. If $\pi_i$ and $\mu_{\Delta y, i}$ can be estimated consistently, a consistent estimator for the true mean change from baseline with return-to-baseline imputation is given by
\begin{equation} \label{eq:est_ML}
\hat \mu_{\Delta \tilde y, i} = \hat \pi_i \hat \mu_{\Delta y, i}.
\end{equation}
Zhang et al. \cite{zhang2020likelihood} obtains the same estimator as (\ref{eq:est_ML}), but lacks clear assumptions for the missingness and clear derivations for the estimator. In addition, the authors provide a variance estimation that assumes  $\hat \pi_i$ and $\hat \mu_{\Delta y, i}$ are independent, which holds only for the case of MCAR \citep{molenberghs1997linear}. Obtaining analytic form for the variance estimation is a challenge, as the correlation between $\hat \pi_i$ and $\hat \mu_{\Delta y, i}$ is generally difficult to estimate if using an existing mixed linear model package to estimate the parameters $\mu_{\Delta y, i}$'s. Therefore, bootstrap is a convenient method to estimate the variance as well as the confidence interval for $\mu_{\Delta \tilde y, i}$ \citep{zhang2020likelihood}. 


\section{Simulations} \label{sec:simulation}
{\color{black} We conducted various simulation experiments to evaluate the performance of the newly proposed return-to-baseline imputation method compared with a traditional return-to-baseline imputation method. Scenarios of having postbaseline measurements at only 1 time point and multiple time points were both examined. We compared how the 2 methods perform under 2 major missing data assumptions: MCAR and MDO. For each imputation method, we also evaluated whether the distribution of the imputed data $\tilde Y_{ijk}$ maintains the same distribution as the original data $Y_{ijk}$ via their mean and standard deviation. Keeping the data distribution of the imputed data the same as that of the original data is important because the imputed data may be used to estimate quantities other than the mean. We also computed the bias, standard deviation, and coverage probabilities of the estimators for common estimand quantities in clinical trials, such as  the mean change from baseline for each treatment arm and the corresponding treatment difference.
}
\subsection{Simulation setting}

We considered a variable with $K+1$ repeated measures $Y_{ij0}, Y_{ij1}, \ldots, Y_{ijK}$ at time $0=t_0 < t_1, \ldots, t_K$. Relating to the notation in Section \ref{sec:methods}, $Y_{ij}=Y_{ijK}$ is the measurement at the final time point, $X_{ij}= Y_{ij0}$ is the baseline measurement of the dependent variable, and $W_{ij}=(Y_{ij1}, Y_{ij2}, \ldots, Y_{ij,K-1})'$ are the measurements at the intermediate time points for the response variable. For subject $j$ in each treatment group $i$, we generated $(Y_{ij0}, Y_{ij1} \ldots, Y_{ijK})'$ independently from a multivariate normal distribution with the mean vector of $\mu_i= (\mu_0, \mu_{i1}, \ldots, \mu_{iK})'$ and a compound-symmetry variance-covariance matrix with standard deviation of $\sigma$ and correlation of $\rho$. 
The probability of postbaseline measures being not missing at time point $k$ conditional on the outcome observed at time point $k-1$ was generated from a Bernoulli distribution with the probability of 
\[
\pi(Y_{ijk}) = \{1 + \exp(\alpha_0 + \alpha_1 Y_{ij,k-1})\}^{-1}, \quad\quad k=1,2,\ldots, K. 
\]
\textcolor{black}{Values of $\alpha_0$ and $\alpha_1$ were selected to target certain proportions of missingness in each treatment arm.} The missing patterns were also set to be monotone, that is, if a measurement was missing at time point $k$, all measurements after this time point were set to be missing. 

In all simulations, we considered $n=100$ subjects in each treatment arm and $I=2$ treatment groups: $i=1$ for placebo and $i=2$ for experimental treatment. For the number of postbaseline time points, we considered 2 scenarios: only 1 postbaseline measurement ($K=1$) and 5 postbaseline measurements ($K=5$). We chose $\mu_i = (0, (i-1)/K, 2(i-1)/K,\ldots, (i-1))'$ and $\sigma = 1.0$. For the correlation, we considered 2 scenarios: $\rho = 0$ and $\rho = 0.5$. For missingness, we considered 2 scenarios: MCAR ($\alpha_1 = 0$) and MDO ($\alpha_1 \ne 0$), each with different proportions of missingness. We chose parameters $\alpha_0$ and $\alpha_1$ resulting in the proportion of missingness to be approximately 20-30\% in all cases. 

{\color{black}To evaluate the bias and the performance of the bootstrap method when variance-covariance structures for the longitudinal outcomes are different between treatments, additional simulations were also conducted. The following parameters were used in the simulation:
\begin{enumerate}
    \item Baseline: $E(X_{ij}) = 0$ and $Var(X_{ij}) = 1$.
    \item Placebo treatment: 
$\mu_{1k} = 0$, $Var(Y_{1jk}) = 1$, and $Corr(Y_{1jk}, Y_{1jk'}) = 0.5$ for $0 \le k' < k \le K$. 
\item Experimental treatment: 
$\mu_{2k} = -k/K$, $Var(Y_{2jk}) = 0.8^2$, and $Corr(Y_{2jk}, Y_{2jk'}) = 0.2$ for $0 \le k' < k \le K$.
\end{enumerate}
}

For each of the above scenarios, 5000 simulated samples were generated. The distribution (mean and standard deviation) of the ``complete data" after imputation as well as the analysis results (using an analysis of covariance model with treatment group as a factor and baseline value of the dependent variable as a covariate) for both the traditional imputation method (TIM) in Zhang et al. \cite{zhang2020missing} and the new imputation method (NIM) {proposed in Section \ref{sec:rtb_mi} [Equation (\ref{eq:rtb_mi})]} were summarized. For each sample, 200 multiple imputation samples were generated. The mean estimate was based on the 200 estimates, and the variance was calculated using the method of combining within- and between-imputation variance \cite{rubin1976inference} and/or a bootstrap method. To evaluate the performance of both methods, the true mean for change from baseline with return-to-baseline imputation was calculated using Equation (\ref{eq:E_ML}). 

\subsection{Simulation results}

\textcolor{black}{The results of the simulation experiments are presented and discussed in this section. Table \ref{table_simu_K1} contains results for scenarios with 1 postbaseline measurement. Table \ref{table_simu_K5} contains results for scenarios with 5 postbaseline measurements. For all simulation scenarios in Table \ref{table_simu_K1} and Table \ref{table_simu_K5}, the choices of $\alpha_0$ and $\alpha_1$ and the corresponding average proportions of missing values in $Y$ at the time point $K$ are reported in Table \ref{table_setting}. A few additional scenarios where each treatment arm has different variance-covariance structure are presented in Table \ref{table_simu_add}, together with the corresponding $\alpha_0$ and $\alpha_1$ values.}

\begin{table}[h!tb] \centering
\caption{The proportion of missing values in $Y_{ijK}$}
\begin{tabular}{cccccc}
\hline\hline
 & \multirow{2}{*}{Missingness} & \multirow{2}{*}{$\rho$}  & \multirow{2}{*}{$(\alpha_0, \alpha_1)$} & \multicolumn{2}{c}{Pr(Missing)} \\ \cline{5-6}
 & &  &  & P  & E  \\ 
\hline \multirow{4}{2cm}{$K = 1$} & \multirow{2}{*}{MCAR}
 & 0 & $(-0.85, 0)$ & 0.300 & 0.300\\
 && $0.5$ & $(-0.85, 0)$ & 0.300 & 0.300 \\
\cline{3-6} & \multirow{2}{*}{MDO}
 & $0$ & $(-1, 1)$ & 0.304 & 0.303 \\
 && $0.5$ & $(-1, 1)$ & 0.304 & 0.303\\
\hline \multirow{4}{2cm}{$K = 5$} & \multirow{2}{*}{MCAR}
 & $0$ & $(-2.60, 0)$ & 0.301& 0.302\\
 && $0.5$ & $(-2.60, 0)$ &0.301 & 0.302\\
\cline{3-6} & \multirow{2}{*}{MDO}
 & $0$ & $(-2.64, 1)$ & 0.390 & 0.300\\
 && $0.5$ & $(-2.57, 1)$ & 0.384 & 0.299\\ \hline\hline
\end{tabular}\\
    {\begin{flushleft} Abbreviations: E, experimental treatment arm; MCAR, missing completely at random; MDO, missingness depending on the observed values; P, placebo arm; Pr(Missing), proportion of missing values for outcome $Y_K$; $\rho$, correlation between baseline and postbaseline measurements; $\alpha_0$,$\alpha_1$, tuning parameters for probability of missing. \end{flushleft} }
\label{table_setting}
\end{table}

\begin{table}[h!tb] \centering
\caption{Summary of the simulation results for the performance of return-to-baseline imputation for cases with only 1 postbaseline measurement
(based on 5000 simulated samples)}
\begin{tabular}{ccccccccccc}
\hline\hline
& & & \multicolumn{2}{c}{$\tilde Y$} && \multicolumn{5}{c}{$\mu_{\Delta \tilde Y}$} \\
\cline{4-5} \cline{7-11}
\rb{Setting} & \rb{Method} & \rb{Group} & Mean & SD && True & Bias & SD & SE & CP  \\ 
\hline \multirow{6}{2cm}{MCAR \\$\rho=0$ } & \multirow{3}{*}{TIM}
 & P	&-0.001	&1.257&&	0.000	&-0.001	&0.107	&0.149	&0.992\\ 
&& E	&-0.701	&1.337&&	-0.700	&0.000	&0.118	&0.149	&0.986\\ 
&& E-P	&     	&     &&	-0.700	&0.000	&0.143	&0.211	&0.997\\ \cline{3-11} & \multirow{3}{*}{NIM}
 & P	&0.000	&1.000&&	0.000	&0.000	&0.096	&0.115	&0.980 \\ 
&& E	&-0.701	&1.099&&	-0.700	&0.000	&0.109	&0.115	&0.962\\ 
&& E-P	&     	&     &&	-0.700	&-0.001	&0.127	&0.164	&0.988\\ \hline  \multirow{6}{2cm}{MCAR \\ $\rho=0.5$} & \multirow{3}{*}{TIM}
 & P	&-0.001	&1.133&&	0.000	&0.000	&0.080	&0.113	&0.993\\ 
&& E	&-0.701	&1.220&&	-0.700	&0.000	&0.093	&0.113	&0.981\\ 
&& E-P	&     	&     &&	-0.700	&0.000	&0.117	&0.160	&0.993\\ \cline{3-11} & \multirow{3}{*}{NIM}
 & P	&-0.001	&0.998&&	0.000	&0.000	&0.077	&0.101	&0.989 \\ 
&& E	&-0.701	&1.096&&	-0.700	&-0.001	&0.091	&0.101	&0.969 \\ 
&& E-P	&     	&     &&	-0.700	&-0.001	&0.115	&0.143	&0.982 \\ \hline \multirow{6}{2cm}{MDO \\ $\rho=0$} & \multirow{3}{*}{TIM}
 & P	&0.177	&1.252&&	0.000	&0.178	&0.104	&0.147	&0.850\\ 
&& E	&-0.518	&1.422&&	-0.697	&0.180	&0.115	&0.147	&0.830\\ 
&& E-P	&     	&     &&	-0.697	&0.001	&0.144	&0.209	&0.996\\  \cline{3-11} & \multirow{3}{*}{NIM}
 & P	&0.001	&1.002&&	0.000	&0.002	&0.100	&0.118	&0.976\\ 
&& E	&-0.696	&1.101&&	-0.697	&0.002	&0.106	&0.118	&0.970\\ 
&& E-P	&     	&     &&	-0.697	&0.000	&0.131	&0.167	&0.988\\  \hline  \multirow{6}{2cm}{MDO \\ $\rho=0.5$} & \multirow{3}{*}{TIM}
 & P	&0.088	&1.144&&	0.000	&0.089	&0.079	&0.113	&0.951\\ 
&& E	&-0.608	&1.347&&	-0.697	&0.090	&0.091	&0.113	&0.927\\ 
&& E-P	&     	&     &&	-0.697	&0.000	&0.118	&0.160	&0.992 \\  \cline{3-11} & \multirow{3}{*}{NIM}
 & P	&0.000	&1.000&&	0.000	&0.001	&0.080	&0.103	&0.986 \\ 
&& E	&-0.698	&1.175&&	-0.697	&0.000	&0.089	&0.103	&0.976\\ 
&& E-P	&     	&     &&	-0.697	&-0.001	&0.118	&0.146	&0.987\\ 
\hline
\hline 
\end{tabular}\\
    {\begin{flushleft} Notations and abbreviations: $\tilde Y$, the ``complete" data for the outcome variable after imputation; $\mu_{\Delta \tilde Y}$, the estimated mean change from baseline in the outcome variable; $\rho$, correlation between baseline and postbaseline measurements; CP, coverage probability of the 95\% confidence interval; E, experimental treatment arm; MCAR, missing completely at random; MDO, missingness depending on the observed values; NIM, new imputation method; P, placebo arm; SD, standard deviation of estimates of the mean; SE, mean standard error estimates of the mean; TIM, traditional imputation method.
    \end{flushleft} }
\label{table_simu_K1}
\end{table}

\begin{table}[h!tb] \centering
\caption{Summary of the simulation results for the performance of return-to-baseline imputation for cases with 5 postbaseline measurements
(based on 5000 simulated samples)}
\begin{tabular}{ccccccccccc}
\hline\hline
& & & \multicolumn{2}{c}{$\tilde Y$} && \multicolumn{5}{c}{$\mu_{\Delta \tilde Y}$} \\
\cline{4-5} \cline{7-11}
\rb{Setting} & \rb{Method} & \rb{Group} & Mean & SD && True & Bias & SD & SE & CP  \\ 
\hline \multirow{6}{2cm}{MCAR \\ $\rho=0$ } & \multirow{3}{*}{TIM}
 & P	&0.001	&1.257	&&0.000		&-0.001	&0.107	&0.149	&0.993\\ 
&& E	&-0.702	&1.341	&&-0.698	&-0.004	&0.116	&0.149	&0.987\\ 
&& E-P	&     	&     	&&-0.698	&-0.004	&0.143	&0.211	&0.996\\ \cline{3-11} & \multirow{3}{*}{NIM}
 & P	&0.001	&1.020	&&0.000		&0.000	&0.098	&0.118	&0.979 \\ 
&& E	&-0.700	&1.121	&&-0.698	&-0.003	&0.108	&0.118	&0.969\\ 
&& E-P	&     	&     	&&-0.698	&-0.003	&0.129	&0.168	&0.987\\ \hline  \multirow{6}{2cm}{MCAR \\ $\rho=0.5$ } & \multirow{3}{*}{TIM}
 & P	&0.002	&1.133	&&0.000		&-0.001	&0.079	&0.113	&0.994\\ 
&& E	&-0.700	&1.224	&&-0.698	&-0.003	&0.092	&0.113	&0.985\\ 
&& E-P	&     	&     	&&-0.698	&-0.002	&0.116	&0.161	&0.993\\ \cline{3-11} & \multirow{3}{*}{NIM}
 & P	&0.002	&1.009	&&0.000		&0.000	&0.076	&0.102	&0.990 \\ 
&& E	&-0.699	&1.111	&&-0.698	&-0.002	&0.089	&0.102	&0.976 \\ 
&& E-P	&     	&     	&&-0.698	&-0.002	&0.113	&0.145	&0.987 \\ \hline \multirow{6}{2cm}{MDO \\ $\rho=0$ } & \multirow{3}{*}{TIM}
 & P	&0.056	&1.329	&&0.000		&0.054	&0.101	&0.156	&0.992\\ 
&& E	&-0.643	&1.376	&&-0.700	&0.056	&0.117	&0.150	&0.978\\ 
&& E-P	&     	&     	&&-0.700	&0.002	&0.144	&0.217	&0.998\\  \cline{3-11} & \multirow{3}{*}{NIM}
 & P	&0.002	&1.040	&&0.000		&0.001	&0.092	&0.122	&0.991\\ 
&& E	&-0.701	&1.123	&&-0.700	&-0.002	&0.107	&0.120	&0.974\\ 
&& E-P	&     	&     	&&-0.700	&-0.003	&0.128	&0.172	&0.990\\  \hline  \multirow{6}{2cm}{MDO \\ $\rho=0.5$} & \multirow{3}{*}{TIM}
 & P	&0.027	&1.173	&&0.000		&0.025	&0.076	&0.118	&0.996\\ 
&& E	&-0.672	&1.326	&&-0.701	&0.027	&0.093	&0.115	&0.981\\ 
&& E-P	&     	&     	&&-0.701	&0.002	&0.118	&0.165	&0.995 \\  \cline{3-11} & \multirow{3}{*}{NIM}
 & P	&0.001	&1.020	&&0.000		&-0.001	&0.076	&0.108	&0.994 \\ 
&& E	&-0.702	&1.200	&&-0.701	&-0.003	&0.091	&0.106	&0.980\\ 
&& E-P	&     	&     	&&-0.701	&-0.002	&0.117	&0.152	&0.989\\ 
\hline
\hline 
\end{tabular}\\
    {\begin{flushleft} Notations and abbreviations: $\tilde Y$, the ``complete" data for the outcome variable after imputation; $\mu_{\Delta \tilde Y}$, the estimated mean change from baseline in the outcome variable; $\rho$, correlation between baseline and postbaseline measurements; CP, coverage probability of the 95\% confidence interval; E, experimental treatment arm; MCAR, missing completely at random; MDO, missingness depending on the observed values; NIM, new imputation method; SD, standard deviation of estimates of the mean; SE, mean standard error estimates of the mean; TIM, traditional imputation method.
    \end{flushleft} }
\label{table_simu_K5}
\end{table}

{\color{black} In Table \ref{table_simu_K1}, \ref{table_simu_K5} and \ref{table_simu_add}, the first 2 columns of summaries are the mean and standard deviation of $\tilde Y_{ijK}$, which is the complete data after imputation was conducted at the last postbaseline time point. The resulting mean and standard deviation are averaged across 5000 simulations. The next 5 columns of the summary data are the true value, the average bias, the standard deviation of the estimates, the estimated standard error, and the coverage probability for the mean change from baseline at each treatment arm and treatment difference based on 5000 simulations. Table \ref{table_simu_add} contains 2 additional columns for estimated standard error and coverage probability using bootstrap.

Presented in Table \ref{table_simu_K1}, scenarios with only 1 postbaseline measurements well illustrate the overall trends that can be observed across various result tables. From the first 2 columns of the result table we can assess whether the new method and traditional method satisfy Criteria C1-C2 in Section \ref{sec:methods} empirically. For the placebo arm, the distributions for $Y$ at baseline and postbaseline were assumed to be the same in the data generating process, therefore we should expect the complete data after the return-to-baseline imputation to have the same distribution as $Y$ (Criteria C1-C2 in Section \ref{sec:methods}). For NIM, the mean and standard deviation of $\tilde Y$ (complete data after imputation) at the \textcolor{black}{last} postbaseline time point \textcolor{black}{($\tilde Y_{ijK}$)} for the placebo arm were close to the true baseline mean (0) and standard deviation (1.0) for all settings. This confirms Criteria C1-C2 are satisfied. For the experimental treatment arms computed by the new imputation method, the sample mean of the ``complete data" was close to the true mean under the return-to-baseline imputation in Equation (\ref{eq:E_ML}). However, the standard deviation was generally larger than the standard deviation of the measurements at baseline or at the final postbaseline time point. For TIM, the mean for the complete data after imputation at the last postbaseline time point was similar compared to baseline for MCAR, but not for MDO. The standard deviation for the postbaseline time point for the placebo arm was always larger than the true standard deviation of 1.0. This shows TIM does not provide a consistent estimator for the distribution of the data under either MCAR and MDO. In all cases, TIM always resulted in larger standard deviation for the ``complete data" after imputation than NIM, which is consistent with the theoretical property that adding an additional error inflates the variance of the distribution of the ``complete" data.} 

{\color{black}The last 5 columns of the Table \ref{table_simu_K1} (the true value, and the average bias, the standard deviation of the estimates, the estimated standard error, and the coverage probability of the 95\% confidence interval for the mean change from baseline at each treatment arm and treatment difference) are inferential statistics calculated based on 5000 simulations. The estimated standard error was estimated using Rubin's rule of combining within- and between-imputation variabilities. The 95\% confidence interval was calculated using the normal approximation based on the estimated standard error.} For the inference for the mean change from baseline to the final time point, NIM resulted in little bias for both MCAR and MDO, while TIM seemed only unbiased for MCAR. For TIM, the bias in each treatment was large (bias between 0.09 and 0.18, relative bias between 13\% and 25\% for the experimental treatment group) for each treatment group under MDO, but the bias in the treatment difference was small. It was likely a coincidence that the biases were canceled out in the simulation setting when both treatment groups had the same variance-covariance structure for the measurements over time. The estimated standard errors based on Rubin's method \cite{rubin1976inference} were larger than the standard deviations of the mean estimates. This phenomenon is consistent with the known fact that Rubin's method generally overestimates the variance when the analysis and imputation models are uncongenial \cite{bartlett2020bootstrap,meng1994multiple,xie2017dissecting}. As a result, the 95\% confidence interval for NIM had a larger coverage probability than the nominal level despite NIM having a minimum bias. 

\begin{table}[h!tb] \centering
\caption{Summary of simulation results for the performance of return-to-baseline imputation with different variance-covariance matrices between treatment groups 
(based on 5000 simulated samples) }
\footnotesize
\begin{tabular}{ccccccccccccc}
\hline\hline
& & & \multicolumn{2}{c}{$\tilde Y$} && \multicolumn{7}{c}{$\mu_{\Delta \tilde Y}$} \\
\cline{4-5} \cline{7-13}
\rb{Setting} & \rb{Method} & \rb{Group} & Mean & SD && True & Bias & SD & SE & CP & $\mbox{SE}_b$ & $\mbox{CP}_b$ \\ 
\hline \multirow{6}{2cm}{$K=1$ \\ MCAR \\ $\alpha_0 = -0.85$ \\ $\alpha_1 = 0$ } & \multirow{3}{*}{TIM}
 & P       & 0     & 1.133 && 0     & 0.000     & 0.081 & 0.116 & 0.995 & 0.080 & 0.946 \\ 
&& E       &-0.701 & 1.155 &&-0.700 & 0.000     & 0.101 & 0.120 & 0.976 & 0.098 & 0.939 \\ 
&& E $-$ P &       &       &&-0.700 & 0.000     & 0.119 & 0.167 & 0.993 & 0.117 & 0.942 \\ 
\cline{3-13} & \multirow{3}{*}{NIM}
 & P       & 0     & 0.999 && 0     & 0.000     & 0.077 & 0.099 & 0.988 & 0.076 & 0.942 \\ 
&& E       &-0.701 & 0.920 &&-0.700 & 0.000     & 0.094 & 0.098 & 0.955 & 0.091 & 0.940 \\ 
&& E $-$ P &       &       &&-0.700 & 0.000     & 0.111 & 0.140 & 0.983 & 0.110 & 0.944 \\  
\hline  \multirow{6}{2cm}{$K=1$ \\ MDO \\ $\alpha_0 = -1.05$ \\ $\alpha_1 = 1$} & \multirow{3}{*}{TIM}
 & P       & 0.088 & 1.141 && 0     & 0.088 & 0.080 & 0.115 & 0.953 & 0.079 & 0.795 \\ 
&& E       &-0.559 & 1.260 &&-0.706 & 0.147 & 0.097 & 0.118 & 0.808 & 0.095 & 0.649 \\ 
&& E $-$ P &       &       &&-0.706 & 0.059 & 0.121 & 0.165 & 0.983 & 0.119 & 0.915 \\ 
\cline{3-13} & \multirow{3}{*}{NIM}
 & P       & 0     & 1.000 && 0     & 0.001 & 0.081 & 0.099 & 0.984 & 0.079 & 0.942 \\ 
&& E       &-0.706 & 0.950 &&-0.706 & 0.000     & 0.088 & 0.098 & 0.967 & 0.086 & 0.942 \\ 
&& E $-$ P &       &       &&-0.706 &-0.001 & 0.112 & 0.139 & 0.985 & 0.111 & 0.946 \\ 
\hline \multirow{6}{2cm}{$K=5$ \\ MCAR \\ $\alpha_0 = -2.6$ \\ $\alpha_1 = 0$} & \multirow{3}{*}{TIM}
 & P       & 0.001 & 1.133 && 0     & 0.000     & 0.081 & 0.116 & 0.996 & 0.080 & 0.945 \\ 
&& E       &-0.698 & 1.159 &&-0.698 &-0.002 & 0.098 & 0.120 & 0.982 & 0.097 & 0.947 \\ 
&& E $-$ P &       &       &&-0.698 &-0.001 & 0.118 & 0.167 & 0.995 & 0.117 & 0.948 \\  
\cline{3-13} & \multirow{3}{*}{NIM}
 & P       & 0     & 1.01  && 0     &-0.001 & 0.077 & 0.100 & 0.991 & 0.075 & 0.946 \\ 
&& E       &-0.698 & 0.935 &&-0.698 &-0.002 & 0.090 & 0.100 & 0.966 & 0.091 & 0.951 \\ 
&& E $-$ P &       &       &&-0.698 &-0.001 & 0.110 & 0.141 & 0.988 & 0.110 & 0.952 \\ 
\hline  \multirow{6}{2cm}{$K=5$ \\ MDO \\ $\alpha_0 = -2.5$ \\ $\alpha_1 = 1$} & \multirow{3}{*}{TIM}
 & P       & 0.025 & 1.179 && 0     & 0.024 & 0.075 & 0.121 & 0.996 & 0.074 & 0.932 \\ 
&& E       &-0.636 & 1.229 &&-0.696 & 0.058 & 0.097 & 0.121 & 0.968 & 0.097 & 0.902 \\ 
&& E $-$ P &       &       &&-0.696 & 0.034 & 0.118 & 0.171 & 0.995 & 0.116 & 0.939 \\ 
\cline{3-13} & \multirow{3}{*}{NIM}
 & P       & 0     & 1.024 && 0     &-0.001 & 0.074 & 0.105 & 0.994 & 0.073 & 0.942 \\ 
&& E       &-0.696 & 0.969 &&-0.696 &-0.002 & 0.089 & 0.101 & 0.974 & 0.090 & 0.954 \\ 
&& E $-$ P &       &       &&-0.696 & 0.000     & 0.111 & 0.146 & 0.990 & 0.111 & 0.948 \\ 
\hline
\hline 
\end{tabular}\\
    {\begin{flushleft} \small Notations and abbreviations: $\tilde Y$, the ``complete" data for the outcome variable after imputation; $\mu_{\Delta \tilde Y}$, the estimated mean change from baseline in the outcome variable; $\rho$, correlation between baseline and postbaseline measurements; CP, coverage probability of the 95\% confidence interval using Rubin's method; $\mbox{CP}_b$, coverage probability of the 95\% confidence interval using bootstrap; E, experimental treatment arm; MCAR, missing completely at random; MDO, missingness depending on the observed values; NIM, new imputation method; P, placebo arm; SD, standard deviation of estimates of the mean; SE, mean standard error estimates of the mean based on Rubin's method; $\mbox{SE}_b$, mean standard error estimates of the mean based on bootstrap; TIM, traditional imputation method.
    \end{flushleft} }
\label{table_simu_add}
\end{table}

Table \ref{table_simu_K5} shows the simulation results for cases with $K=5$ postbaseline measurements. \textcolor{black}{Results were generally similar to those for $K=1$}. For the placebo arm, the standard deviation for the data after imputation at the final time point was always larger compared to the baseline value for TIM. Again, for the case of MDO, NIM had little bias in the estimator for the change from baseline in $Y$, while the traditional showed a clear bias; however, the bias for each treatment group was smaller (bias between 0.027 and 0.056, relative bias approximately 4\% and 8\% for the experimental treatment group) compared to the cases of $K=1$. This is because the probability of missing values only depends on the observed value at the previous measurement and the correlation between the missing value indicator for the final visit and the baseline value becomes much smaller for $K=5$ compared to $K=1$. In this case, the missingness is closer to MCAR compared to the case of $K=1$. In practice, the probability of missingness for all postbaseline time points may depend on the baseline value even after conditioning on the observed postbaseline values, in which case the bias could be larger.

Additional simulations were performed and summarized in Table 4 to assess the bias when variance-covariance structures for the longitudinal outcomes
are different between treatments and performance of bootstrap method. For each simulated sample in Table \ref{table_simu_add}, 100 bootstrap samples were generated, and the variance of the 100 bootstrap estimates was used to estimate the variances of estimators for the mean change from baseline in $Y$ for each treatment group and the treatment difference. The 95\% bootstrap confidence intervals were constructed using normal approximation. To improve the coverage probability, we made a small adjustment based on $t$-distribution. Essentially, the $100(1-\alpha)$ bootstrap confidence interval was constructed as $\hat \theta \pm t_{1-\alpha'/2, \tilde n-2} \hat \sigma$, where $t_{1-\alpha'/2,\tilde n-2}$ is the $1-\alpha'/2$ quantile of the $t$-distribution with degrees of freedom equal to $\tilde n-2$, $\tilde n$ is the number of subjects without missing values, $\hat \sigma$ is the bootstrap standard error, $\alpha'$ is calculated from $\alpha'/2 = \Phi(\tilde n/(\tilde n-2) t_{\alpha/2,\tilde n-2})$, and $\Phi$ is the cumulative distribution function of the standard normal distribution. This adjustment is borrowed from the idea of the expanded percentile bootstrap confidence interval \cite{hesterberg2015teachers}. Table 4 shows a summary of simulation results. For MDO, TIM produced biased mean estimates for each treatment group and the treatment difference. For NIM, there was little bias in all cases. In all cases, the bootstrap standard errors for the new method were very close {\color{black}to} the standard deviation of the estimates of 5000 simulated samples and the 95\% bootstrap confidence intervals always had a coverage probability of approximately 0.95. 

\section{Summary and Conclusion}
Return-to-baseline imputation is an important imputation method to estimate potential outcomes assuming the mean response after certain intercurrent events is the same as baseline. For disease with very slow progression, when the treatment does not have a disease modification effect and there is little placebo effect, one can assume the response may return to baseline after the discontinuation of the study medication. One plausible example is the hemoglobin A1c (HbA1c) in diabetes disease area. The disease progression is slow, most anti-diabetes treatments do not have a disease modification effect, and there is little placebo effect as HbA1c is an objective measurement from blood samples. It can also serve as a sensitivity analysis for handling missing values under a conservative assumption that the expected value of the missing measurement is the same as at baseline. Analyses based on return-to-baseline imputation are often requested by regulatory agencies either as the primary analysis or sensitivity analysis. The current approaches for implementing return-to-baseline imputation in literature and practice always inflate the variance of the data after imputation. In addition, they result in biased mean for ``complete" data after imputation when the probability of missingness depends on the observed baseline or postbaseline intermediate outcomes. We provide a theoretical framework describing the required statistical criteria for a return-to-baseline imputation method. Under this framework, we propose a new method for return-to-baseline multiple imputation. The new method can be easily implemented using existing multiple imputation procedures in statistical analysis packages such as R or SAS. Simulation studies show the new method provides a consistent estimator and outperforms existing methods in terms of bias and variance for certain scenarios.

In the theoretical derivations in Section \ref{sec:methods}, for convenience, we first assumed the baseline variable $X_{ij}$ and ancillary variables $Z_{ij}$ were not missing; however, in practice, for longitudinal data, the new method is still valid when some variables in $Z_{ij}$ are missing as long as the missingness does not depend on unobserved outcomes. Simulation results in Tables \ref{table_simu_K1}, \ref{table_simu_K5}, and \ref{table_simu_add} show the new method produces consistent estimators when intermediate outcomes are missing at random. 

Inclusion and exclusion criteria for the clinical study may truncate the baseline variable $X$. The return-to-baseline imputation methods based on multivariate normal distributions may produce imputed values outside of the required range for $X$ based on the inclusion criteria. One approach is to truncate the data after the imputation. Another approach is to use the predictive mean matching (PMM) method, which generates imputed values from neighborhood observed baseline values \cite{rubin1986statistical,little1988missing}. {\color{black}However, the PMM method cannot be directly implemented with the imputation package using the PMM option.} It has to be done as an additional step by comparing the imputed values in Equations (\ref{eq:return_to_baseline_imputation}) or (\ref{eq:rtb_imputation2}) with the baseline observations, and randomly draw a number from the closest $N_p$ observation, where $N_p$ is a predefined integer number.

For the newly proposed method, we ``borrow" the correlations between $X_{ij}$ and $Y_{ij}$ in constructing the conditional distribution (\ref{eq:return_to_baseline_imputation}). A potential counter-argument would be the correlation of $X_{ij}$ and \textcolor{black}{$Y_{ij}^*$} (a repeated measure of $X_{ij}$ under the same distribution) would not be the same as $Corr(X_{ij},Y_{ij})$. Note, this is also true for the traditional method (the imputed value is obtained by adding an independent random error to the baseline value) which {\color{black}implicitly} uses a specific correlation and arbitrarily inflates the variance [variance for the imputed value neither equal to $Var(X_{ij})$ nor equal to $Var(Y_{ij})$] during the imputation. In addition, it is difficult to define \textcolor{black}{$Corr(X_{ij}, Y_{ij}^*)$}: 2 repeated measures within a shorter time interval may have a different correlation compared to 2 repeated measures within a longer time interval (e.g, considering a time series model) even if the mean does not change over time. Similarly, there is no strong argument that $Var(Y_{ij}^*)$ must be equal to $Var(X_{ij})$, as the variability may increase over time even when the mean remains the same as baseline. Therefore, there are at least 2 versions of return-to-baseline imputation based on the assumption for the variance of the imputed value: only ``return" the mean to baseline and ``return" both the mean and variance to baseline. In this article, we have predominately studied the former case; however, the latter version of return-to-baseline imputation is given in Equation (\ref{eq:rtb_imputation2}) as 1 additional option. 

The {\color{black}proposed} return-to-baseline imputation method has limitations and requires careful considerations for application. First, the hypothetical return-to-baseline potential outcome is only plausible for diseases with slow progression and for treatments without disease modification. For example, if a disease has rapid progression (e.g, tumor growth), assuming patients with certain intercurrent events will have postbaseline measurements similar to baseline does not sound reasonable. Second, the new return-to-baseline imputation method assumes the baseline value is not missing and the probability of missingness does not depend on the unobserved values. We also assume $(X, Z', Y)'$ is from a multivariate normal distribution. In practice, it is common to have scenarios where $Y$ is continuous but $Z$ is binary or {\color{black}categorical}. In this case, we may still approximately consider $(X, Z', Y)'$ a multivariate normal distribution, because the major interest is to impute $Y$ and the misspecification for $Z$ is less critical. However, in cases where the outcome variable $Y$ is binary, it would be more complex. If the binary endpoint is derived from a continuous outcome, we can use the newly proposed method to impute the continuous outcome and then derive the binary endpoint. The imputation procedure will be more complex for non-Gaussian distributions, which is a challenge for all multiple imputation based approaches. For a binary endpoint without a underlying continuous variable, direct imputation of the return-to-baseline value will be more complex and require additional research, \textcolor{black}{similar to the imputation of missing data for a binary variable under ignorable missingness \cite{lipkovich2005multiple, floden2019imputation,
grobler2020multiple,
ma2021analysis}.} More details on multiple imputation can be found in Van Buuren \cite{van2018flexible}. 

In conclusion, we propose a new method for the return-to-baseline imputation that outperforms existing methods and can be easily implemented with the existing multiple imputation packages.  
This new method may be used for analysing clinical trial data when the return-to-baseline imputation is needed.

\section*{Acknowledgements}
We would like to thank Ilya Lipkovich and Govinda Weerakkody for their useful comments and their careful review of this manuscript, and Dana Schamberger for an editorial review of the manuscript. 
\bibliographystyle{vancouver}
\bibliography{references}

\end{document}